\documentstyle[12pt,epsf]{article}
\newcommand{\beq}{\begin{equation}}
\newcommand{\eeq}{\end{equation}}
\newcommand{\beqa}{\begin{eqnarray}}
\newcommand{\eeqa}{\end{eqnarray}}

\renewcommand{\arraystretch}{1.1}
\newcommand{\vs}{\vspace{-0.25cm}}

\begin{document}

{\small \hfill FZJ-IKP(TH)-1999-07}


\vspace{2cm}

\begin{center}

{{\Large \bf 
Baryon form factors}}

\end{center}

\vspace{.3in}

\begin{center}
{\large
B. Kubis\footnote{email: b.kubis@fz-juelich.de}, 
T.R. Hemmert\footnote{email: th.hemmert@fz-juelich.de}, 
Ulf-G. Mei{\ss}ner\footnote{email: Ulf-G.Meissner@fz-juelich.de}}

\bigskip

{\it Forschungszentrum J\"ulich, Institut f\"ur Kernphysik
(Theorie)\\ D-52425 J\"ulich, Germany}

\bigskip

\end{center}

\vspace{.7in}

\thispagestyle{empty}

\begin{abstract}
\noindent
We calculate the form factors of the baryon octet in the
framework of heavy baryon chiral perturbation theory.
The calculated charge radius of the $\Sigma^-$ is in agreement 
with recent measurements from CERN and Fermilab. 
We show that kaon loop effects can play a significant role
in the neutron electric form factor. Furthermore, we derive generalized
Caldi--Pagels relations between various charge radii which are 
free of chiral loop  effects. 
\end{abstract}

\vspace{1in}


\vfill

\pagebreak


\noindent {\bf 1.} Hadrons are composite objects which are characterized
by certain sizes. The latter depend on the type of probe which is
used to investigate the hadron structure. In the realm of strong QCD,
the most precise information can be obtained using the electromagnetic (em)
current. One measures the so--called electric and magnetic form factors,
which in the Breit frame can be interpreted as the distribution of electric
charge and magnetization within the hadron. The em form factors of the
nucleon have been mapped out to high precision using electron scattering 
off protons and deuterons  over many decades, for
a comprehensive reference see e.g.~\cite{mmd}. The situation is very
different for the other members of the ground state baryon octet, the hyperons.
Until recently, for these only the magnetic moments had been
determined. Last year, the first measurement
of the $\Sigma^-$ charge radius has been reported from experiments
performed at CERN (WA89)~\cite{wa89} and FNAL (E781)~\cite{selex} using 
electron 
scattering in inverse kinematics. More precisely, a very high energetic
hyperon beam was scattered off atomic shell electrons and from the
momentum dependence of the cross section the form factor could be extracted.
For example, the SELEX collaboration~\cite{selex} deduced from their data in the
$q^2$--region of -0.03 to -0.16~GeV$^2$ the mean square charge radius
to be
\beq
\langle r^2\rangle_{\Sigma^-}= 0.60 \pm 0.08_{\rm stat.} \pm 0.08_{\rm syst.}
\,\, {\rm fm}^2~.
\eeq
Indeed, from naive quark model considerations one expects the hyperon
size to be smaller than the one of the nucleon because the strange quark
is more massive than the up and down quarks. A systematic investigation
of the hyperon radii would allow one to study the deviations from
flavor SU(3) as it can be done for the magnetic moments. Besides from
these fundamental aspects, a knowledge of the hyperon form factors
is also needed to interpret the kaon electroproduction data which are
and will be obtained
at ELSA (Bonn), MAMI-C (Mainz) and TJNAF. In fact, one can reverse the argument and
try to use the processes $\gamma^\star
+ N \to K +Y$, where $\gamma^\star \, (Y)$ denotes a virtual photon (hyperon),
to determine the hyperon form factors (for a recent discussion, see 
ref.~\cite{bm}). A related issue of current interest is the question
of strangeness in the nucleon, which can either be inferred indirectly from
kaon cloud contributions to the em nucleon form factors or more directly
from a determination of the so--called strange nucleon form factors. 
It has been argued since a long time that in particular the neutron
electric form factor might be particularly sensitive to strange quark
effects, for an early reference see e.g.~\cite{gk}. Furthermore,  
the strange magnetic ff  was recently investigated using chiral 
perturbation theory and  a leading one loop prediction free of adjustable
parameters could be made~\cite{hms}. Here, we use conventional three
flavor baryon chiral perturbation theory in the heavy fermion formalism
(called HBCHPT) to calculate the em form factors of the hyperons and the
strangeness contributions in the nucleon. The momentum
dependence of the ffs is governed by pion and kaon cloud contributions
togther with some counterterms which parametrize the effects of higher
mass states. These effects can be calculated systematically and
precisely. From previous studies of the
nucleon in HBCHPT and extensions thereof~\cite{bfhm}, it is known that
this approach is limited to low momentum transfer, say $-0.2 \le q^2 
\le 0\,$GeV$^2$. Therefore, we will restrict our investigation to this
region of momentum transfer. In the SU(3) study presented here, 
all appearing low--energy
constants can be determined from the octet magnetic moments and the
proton and neutron electric charge radii. Therefore, the momentum dependence
of the em hyperon form factors can be predicted. Of course, many models
have been used to calculate these form factors with wildly varying
predictions, e.g. $0.3\,{\rm fm}^2 \le \langle r^2\rangle_{\Sigma^-}\le 
1.2\,$fm$^2$,  see e.g. fig.3 in~\cite{selex}.
The results obtained in HBCHPT allow one to further constrain these
models.

\medskip

\noindent {\bf 2.}
The starting point of  heavy baryon 
CHPT  is an effective Lagrangian formulated in terms of the asymptotic fields,
here the octet of Goldstone bosons and the ground state baryon octet $B$. 
The Lagrangian admits a low energy expansion of the form
\beq
{\cal L}_{\rm eff} = {\cal L}_M + {\cal L}_{MB} =
{\cal L}_M^{(2)}  +
{\cal L}_{MB}^{(1)} + {\cal L}_{MB}^{(2)}
+ {\cal L}_{MB}^{(3)}  + \ldots \,
\eeq  
where the subscript $'M'$ ($'MB'$) denotes the meson (meson--baryon)
sector and the superscript $'(i)'$ the chiral dimension, which counts
the power  of external momenta and/or meson mass insertions. We call the
corresponding small expansion parameter $q$.
The Goldstone bosons are collected in the familiar SU(3)--valued
field $U = \exp\{i \phi / F_\phi \}$, with $F_\phi$ the octet decay
constant. If not otherwise specified, we set $F_\phi = (F_\pi + F_K)/2
=100\,$MeV. The effects due to $F_K \ne F_\pi$ are of higher order
and can thus be neglected here.
The baryons are  given by a $3\times3$ matrix, transforming 
under SU(3)$_L \times$ SU(3)$_R$ as 
usual matter fields,  $B \to B' = KBK^\dagger$, with $K(U,L,R)$ the 
compensator field which is an element of the conserved subgroup
SU(3)$_V$. We use the notation of \cite{mm}.  The ellipsis
stands for terms  not needed here.  Beyond leading
order, the effective Lagrangian contains parameters not fixed by
chiral symmetry, the so--called low--energy constants (LECs). 
These LECs must be be pinned down from data. 
In what follows, we will work to order $q^3$ 
in the chiral expansion. More precisely, we need the following terms
from the second and third order meson--baryon Lagrangian (we omit
finite terms contributing to the mass shift and alike):
\beqa
{\cal L}_{MB}^{(2)} &=& 
-\frac{i}{4m}  \, \langle \bar{B} [S^\mu,
S^\nu][ F_{\mu \nu}^+, B] \rangle   
-\frac{i\, b_{D/F}}{4m}  \, \langle \bar{B} [S^\mu,
S^\nu] [ F_{\mu \nu}^+, B]_\pm \rangle~, \\
{\cal L}_{MB}^{(3)} &=& \frac{b_{D/F}}{8m^2}  \, \langle \bar{B}
[[v^\mu D^\nu, F_{\mu\nu}^+ ], B]_\pm \rangle -
\frac{d_{101/102}}{(4\pi F_\phi)^2}  \, \langle \bar{B}
[[v^\mu D^\nu, F_{\mu\nu}^+ ], B]_\mp \rangle \nonumber \\
&+& \frac{i d_{33/34}}{(4\pi F_\phi)^2}\biggl(  \, \langle \bar{B}
[\chi_+ , [v\cdot D,B]]_\pm \rangle +{\rm h.c}\biggr)
+ \frac{i d_{35}}{(4\pi F_\phi)^2} \biggl( \, \langle \bar{B}
[v\cdot D,B] \rangle \langle \chi_+ \rangle +{\rm h.c}
\biggr) \nonumber \\
&+& \frac{i d_{27}}{(4\pi F_\phi)^2}  \, \langle \bar{B}
[v\cdot D,[v\cdot D, [v\cdot D,B]]]\rangle~,
\eeqa
with $S_\mu$ the covariant spin--operator, $F_{\mu \nu}^+ =
e(u^\dagger Q F_{\mu\nu}u+ uQF_{\mu\nu}u^\dagger )
= 2e(\partial_\mu A_\nu-\partial_\nu A_\mu)Q + {\cal O}(\phi^2)$ 
and $\langle
\ldots \rangle$ denotes the trace in flavor space. 
Here, $Q={\rm diag}(2,-1,-1)/3$ is the quark charge matrix and 
$u = \sqrt{U}$. Throughout we work in the isospin limit
$m_u = m_d$ and thus neglect the small $\Lambda-\Sigma^0$ mixing.
The dimension two terms obviously give the magnetic coupling of the photon
to the baryons. The corresponding LECs are determined with respect
to the anomalous magnetic moments of the baryons. The third order
LECs $d_{101,102}$ feature prominently in the electric radii whereas
the last terms are only needed for renormalization and thus do not
lead to observable effects.

\medskip

\noindent {\bf 3.} In HBCHPT, the natural frame
 to work with is the Breit--frame~\cite{bkkm}
and thus one defines the electromagnetic Sachs form factors
$G_E (q^2)$ and $G_M(q^2)$. For a given baryon state from the
octet, the matrix element of the electromagnetic current
expressed in terms of the three light quark fields $q^T =(u,d,s)$,
$V_\mu = \bar{q} Q \gamma_\mu q$, takes the form 
(to avoid unnecessary flavor indices, we work in the
physical basis and consider mostly diagonal matrix elements, thus
we can omit labelling the initial and final baryon states; the
exception being the $\Lambda-\Sigma^0$ transition) 
\beq
 \langle B|V_\mu|B \rangle = 
\frac{1}{N_{i}N_{f}} {\bar u(p')} \, P^{+}_{v} \,\left[
G_{E} (q^{2}) v_\mu
+ \frac{1}{m} G_{M} (q^{2}) \left[ S_\mu,S_\nu \right]
q^{\nu}
\right] \, P^{+}_{v}\, u(p)~, 
\eeq 
with
\beq
q_\mu = (p' -p)_\mu~, \quad N =\sqrt{\frac{E+m}{2m}}~,
\eeq
and the $P_v^+$ are positive--velocity projection operators. For a
more detailed discussion of this expression and the relation
to the standard Dirac and Pauli form factors, see e.g.~\cite{bfhm}.
The chiral expansion of the form factors takes the form (here,
$G$ is a genuine symbol for any form factor)
\beq
G (q^2) = G^{\rm tree}(q^2)  + G^{\rm ct}(q^2)  +  G^{\rm loop}(q^2)~,
\eeq
where we have split the tree graphs into the terms with fixed coefficients
and the counterterms. For the magnetic form factors and the electric
ones of the neutral octet baryons, the chiral
expansion starts at second order. We thus
expect to achieve the most precise description of the electric form
factors  for the charged particles.
To third order, we have to deal with four non--trivial one loop
graphs, two of the tadpole  and two of the self--energy type, see 
e.g. ref.~\cite{bfhm}.
In the electric form factors, divergences appear. These are
canceled by the appropriate $\beta$--functions of the operators
101 and 102 from table~1 of ref.\cite{mm},
$\beta_{101} = -(9+25D^2+45F^2)/36$, $\beta_{102} = -5DF/2$.
The general structure of all Sachs\ form factors is 
\beqa\label{GE}
G_E(q^2)&=&Q+\frac{1}{\left( 4\pi F_{\phi }\right) ^{2}}\biggl\{
\alpha_0 + \sum_{X=\pi,K}  \alpha_X \ln\frac{M_X}{\lambda} + 
\sum_{X=\pi,K}\biggl[ -\beta_X \left(2M_X^2-\frac{5}{4}q^2\right) 
\nonumber \\
&+& \gamma_X \left( M_X^2 - \frac{1}{4}q^2\right) \biggr]\, I_E^X(q^2)
-\bigl( 2Qd_{101}^r (\lambda)+\alpha_D d_{102}^r (\lambda)
\bigr)q^2 \biggr\} +{q^2\over
4m^2}\left( Qb_F + \alpha_D b^D\right)~,
\\
G_{M}( q^{2})  &=&Q\left( 1+b_{F}\right) +\alpha _{D}b_{D}
+\frac{m}{16\pi F_{\phi }^{2}}\,\sum_{X=\pi,K} \beta_X
\left[M_X+(M_X^2-\frac{1}{4}q^2) \, I_M^X(q^2)\right]~,
\eeqa
with
\beq
I_E^X (q^2) = \frac{1}{3} \int_0^1 dx \ln \left( 1 -
x(1-x)\frac{q^2}{M_X^2}\right)~,
\quad I_M^X (q^2) = \int_0^1 dx \frac{1}{\sqrt{M_X^2-x(1-x)q^2}}~,
\eeq
and $\lambda$ is the scale of dimensional regularization. Throughout,
we set $\lambda =1\,$GeV.
The corresponding squared electric radii  and magnetic 
slopes\footnote{We work with these slopes instead of the magnetic
radii to avoid the uncertainty arising from the description of the
magnetic moments to this order.} are given by 
\beqa
 r_{E}^{2} = \frac{6}{Q+\delta_{Q0}}\frac{dG_E(q^2)}{dq^2}\biggl|_{q^2=0}
&=&\frac{1}{\left( 4\pi F_{\phi }\right) ^{2}}
\biggl\{ \eta +6\alpha _{\pi }\log \left( \frac{M_{\pi }}{\lambda }\right)
+6\alpha _{K}\log \left( \frac{M_{K}}{\lambda }\right)  \nonumber \\
&& -12\bigl( Qd_{101}^r (\lambda)+
\alpha _{D}d_{102}^r (\lambda)\bigr) \biggr\} 
+\frac{3}{2m^{2}}\left( Qb^{F}+\alpha _{D}b^{D}\right) \\
G_M'(0) = \frac{dG_M(q^2)}{dq^2}\biggl|_{q^2=0}
&=&-\frac{1}{96\pi F_{\phi }^{2}}\left\{ \beta
_{\pi }\frac{m}{M_{\pi }}+\beta _{K}\frac{m}{M_{K}}\right\}
\eeqa
where $\eta =6\alpha _{0}+\frac{2}{3}\beta _{\pi }-\frac{1}{3}\gamma _{\pi
}+\frac{2}{3}\beta _{K}-\frac{1}{3}\gamma _{K}$. Note that the normaliztion of
the electric radii is such that one divides by the charge for a charged
particle, i.e. by $G(0) = Q$, and by one for the neutrals.  The magnetic
moments $\mu \equiv G_M (0)$ are 
\beq\label{mm}
\mu = Q +\kappa =
Q\left( 1+b^{F}\right) +\alpha _{D}b^{D}+\beta _{\pi }\frac{mM_{\pi }}{
8\pi F_{\phi }^{2}}+\beta _{K}\frac{mM_{K}}{8\pi F_{\phi }^{2}}~, 
\eeq
with $\kappa$ the corresponding anaomalous magentic moment.
The pertinent coefficients are shown in the table~1.
\begin{eqnarray*}
&& 
\begin{tabular}{|c|c|c|c|c|}
\hline
& $Q$ & $\alpha _{0}$ & $\alpha _{\pi }$ & $\alpha _{K}$ 
\\ \hline
$p$ & $1$ & 
$-\frac{1}{12}-\frac{85}{108}D^{2}-\frac{17}{18}DF-\frac{17}{12}F^{2}$ & 
$-\frac{1}{6}-\frac{5}{6}\left( D+F\right) ^{2}$ & 
$-\frac{1}{3}-\frac{5}{3}\left( \frac{D^{2}}{3}+F^{2}\right) $ 
\\ 
$n$ & $0$ & 
$\frac{17}{9}DF$ & 
$\frac{1}{6}+\frac{5}{6}\left( D+F\right)^{2} $ & 
$-\frac{1}{6}-\frac{5}{6}\left( D-F\right) ^{2}$ 
\\ 
$\Sigma ^{+}$ & $1$ & 
$-\frac{1}{12}-\frac{85}{108}D^{2}-\frac{17}{18}DF-\frac{17}{12}F^{2}$ & 
$-\frac{1}{3}-\frac{5}{3}\left( \frac{D^{2}}{3}+F^{2}\right) $ & 
$-\frac{1}{6}-\frac{5}{6}\left( D+F\right) ^{2}$ 
\\ 
$\Sigma ^{-}$ & $-1$ & 
$\frac{1}{12}+\frac{85}{108}D^{2}-\frac{17}{18}DF+\frac{17}{12}F^{2}$ & 
$\frac{1}{3}+\frac{5}{3}\left( \frac{D^{2}}{3}+F^{2}\right) $ & 
$\frac{1}{6}+\frac{5}{6}\left( D-F\right) ^{2}$ 
\\ 
$\Sigma ^{0}$ & $0$ & 
$-\frac{17}{18}DF$ & 
$0$ & 
$-\frac{5}{3}DF$ 
\\ 
$\Lambda $ & $0$ & 
$\frac{17}{18}DF$ & 
$0$ & 
$\frac{5}{3}DF$ 
\\ 
$\Lambda \Sigma^{0}$ & $0$ &
$-\frac{17}{6\sqrt{3}}DF$ &
$ -\frac{10}{3\sqrt{3}}DF$ &
$ -\frac{5}{3\sqrt{3}}DF$
\\
$\Xi ^{0}$ & $0$ & 
$\frac{17}{9}DF$ & 
$-\frac{1}{6}-\frac{5}{6}\left(D-F\right) ^{2}$ & 
$\frac{1}{6}+\frac{5}{6}\left( D+F\right) ^{2}$ 
\\ 
$\Xi ^{-}$ & $-1$ & 
$\frac{1}{12}+\frac{85}{108}D^{2}-\frac{17}{18}DF+\frac{17}{12}F^{2}$ & 
$\frac{1}{6}+\frac{5}{6}\left( D-F\right) ^{2}$ & 
$\frac{1}{3}+\frac{5}{3}\left( \frac{D^{2}}{3}+F^{2}\right) $ 
\\ \hline
\end{tabular}
\\
&& 
\begin{tabular}{|c|c|c|c|c|c|c|}
\hline
& $\alpha _{D}$ & $\beta _{\pi }$ & $\gamma_{\pi }$ & $\beta _{K}$ & 
$\gamma_{K}$ & $\eta $ 
\\ \hline
$p$ & $\frac{1}{3}$ & 
$-\left( D+F\right) ^{2}$ & $1$ & 
$-2\left( \frac{D^{2}}{3}+F^{2}\right) $ & $2$ & 
$-\frac{3}{2}-\frac{35}{6}D^{2}-7DF-\frac{21}{2}F^{2}$ 
\\ 
$n$ & $-\frac{2}{3}$ & 
$\left( D+F\right) ^{2}$ & $-1$ & 
$-\left( D-F\right)^{2}$ & $1$ & 
$14DF$ 
\\ 
$\Sigma ^{+}$ & $\frac{1}{3}$ & 
$-2\left( \frac{D^{2}}{3}+F^{2}\right) $ & $2 $ & 
$-\left( D+F\right) ^{2}$ & $1$ & 
$-\frac{3}{2}-\frac{35}{6}D^{2}-7DF-\frac{21}{2}F^{2}$ 
\\ 
$\Sigma ^{-}$ & $\frac{1}{3}$ & 
$2\left( \frac{D^{2}}{3}+F^{2}\right) $ & $-2 $ & 
$\left( D-F\right) ^{2}$ & $-1$ & 
$\frac{3}{2}+\frac{35}{6}D^{2}-7DF+\frac{21}{2}F^{2}$ 
\\ 
$\Sigma ^{0}$ & $\frac{1}{3}$ & 
$0$ & $0$ & $-2DF$ & $0$ & $-7DF$ 
\\ 
$\Lambda $ & $-\frac{1}{3}$ & 
$0$ & $0$ & $2DF$ & $0$ & $7DF$ 
\\ 
$\Lambda \Sigma^{0}$ & $\frac{1}{\sqrt{3}}$ &
$-\frac{4}{\sqrt{3}}DF $ & $0$ &
$-\frac{2}{\sqrt{3}}DF $ & $0$ &
$-7\sqrt{3}DF$
\\
$\Xi ^{0}$ & $-\frac{2}{3}$ & 
$-\left( D-F\right) ^{2}$ & $1$ & 
$\left(D+F\right) ^{2}$ & $-1$ & 
$14DF$ 
\\ 
$\Xi ^{-}$ & $\frac{1}{3}$ & 
$\left( D-F\right) ^{2}$ & $-1$ & 
$2\left(\frac{D^{2}}{3}+F^{2}\right) $ & $-2$ & 
$\frac{3}{2}+\frac{35}{6}D^{2}-7DF+\frac{21}{2}F^{2}$ 
\\ \hline
\end{tabular} 
\end{eqnarray*}
\smallskip
\centerline{Table~1: Table of coefficients for the various octet states.}

\bigskip

\noindent We remark that the last term in eq.(\ref{GE}) can be expressed
in terms of the anomalous magnetic moment of the baryon considered
by use of eq.(\ref{mm}), i.e. it is the well--known Foldy term.
In the limit $M_K \to \infty$, we recover the SU(2) results for the
form factors as demanded by decoupling.
It is worth to notice that we can derive three relations
for the magnetic moments, electric and magnetic charge radii
which are  free of chiral loop effects (these have been found
originally for the magnetic moments only in ref.\cite{cp}). Denoting
by $O$ the observables $r_E^2$, $r_M^2$ and $\mu$, these generalized
Caldi--Pagels relations take the form
\beqa
\frac{1}{2} O \left(\Sigma^+ + \Sigma^- \right)
&=& O  \left(\Sigma^0 \right) = -O  \left(\Lambda \right)~, \\
O\left(\Xi^0 + \Xi^- + p + n\right) &=& 2 O  \left(\Lambda \right)~,\\ 
O\left(\Xi^0 + n + \sqrt{3}\Lambda\Sigma^0\right) &=& O\left(\Lambda \right)~.
\eeqa
The first of these is nothing but isospin symmetry. Note that for
the magnetic moments, these relations work astonishingly well, 
within a few percent.
\medskip

\noindent{\bf 4.} Before presenting results, we must fix parameters.
Throughout, we use $D=3/4$, $F=1/2$ and $F_\phi = 100\,$MeV.
The other parameters are varied as follows
\beqa
{\rm Set~1} &:& b_D = 3.92~, b_F=2.92~,d_{101}^r(1~{\rm GeV})=-1.06~,
d_{102}^r(1~{\rm GeV})=1.70~,
m=0.94\,{\rm GeV}~,
\nonumber \\
{\rm Set~2} &:& b_D = 5.17~, b_F=2.76~,d_{101}^r(1~{\rm GeV})=-1.06~,
d_{102}^r(1~{\rm GeV})=1.70~,
m=0.94\,{\rm GeV}~,
\nonumber \\
{\rm Set~3} &:& b_D = 5.81~, b_F=3.22~,d_{101}^r(1~{\rm GeV})=-1.27~,
d_{102}^r(1~{\rm GeV})=1.68~,
m=1.15\,{\rm GeV}~. \nonumber \\ &&
\eeqa
Here, the first set is chosen such that the proton and neutron electric
radii and magnetic moments are exactly reproduced. In set 2, an
overall best fit to the octet magnetic moments is achieved without
readjusting the LECs $d_{101,102}$. In the third set, the average
octet mass is used while $b_{D,F}$ are obtained from the octet
magnetic moments and $d_{101,102}$ are adjusted to reproduce $r_{E,p}^2$
and $r_{E,n}^2$. We note that for all sets the LECs
$d_{101,102}^r (\lambda=1~{\rm GeV})$ are of natural size.
 While sets~1,3 lead to the same octet charge radii,
the sets~1,2 obviously give the same magnetic slopes. The resulting 
charge radii and magnetic slopes are collected in table~2.
\addtocounter{table}{+1}
\renewcommand{\arraystretch}{1.1}
\begin{table}[htb]
\begin{center}
\begin{tabular}{|l|c|c|}
    \hline
\end{tabular}
\begin{tabular}{|l||r|r|c||r|r|c|}
\hline
%
& \multicolumn{3}{c||}{$r_E^2$ [fm$^2$]}&
\multicolumn{3}{c|}{$G_M' (0)$ [fm$^2$]} \\
\cline{2-7}  B & Set 1 & Set 2 & Exp. & Set 1 & Set 3 & Exp. \\
    \hline
    p                 &  0.735   &  0.717   &  0.735~\cite{mmd}
                      &  0.157   &  0.192   &  0.325~\cite{mmd}\\    
    n                 & $-$0.113 & $-$0.168 & $-$0.113$\pm$0.004~\cite{kop}
                      & $-$0.134 & $-$0.164  & 0.252~\cite{mmd} \\
    $\Sigma^+$        &  0.642   &  0.659  & -- &  0.114 &  0.140 & -- \\
    $\Sigma^-$        &  0.803   &  0.765   
                      & 0.60$\pm$0.08$\pm$0.08~\cite{selex}     
                      & $-$0.077  & $-$0.095   & -- \\
                      &     &     & 0.91$\pm$0.32$\pm$0.40~\cite{wa89}& && \\
    $\Sigma^0$        & $-$0.081 &$-$0.053 & --  & 0.018 & 0.023& --   \\
    $\Lambda$         & 0.081 & 0.053 & --  & $-$0.018 & $-$0.023& --   \\
    $\Xi^0$           & 0.202 & 0.147 &--& $-$0.033 & $-$0.040 &--  \\
    $\Xi^-$           & 0.645 & 0.608 &--& $-$0.027 & $-$0.033 &--   \\
    $\Lambda\Sigma^0$ & $-$0.005 & 0.004 & -- &0.086 &0.150&-- \\
    \hline
  \end{tabular}
\caption{Electric charge radii and magnetic slopes for the octet baryons
for the parameter sets described in the text.}
\end{center}
\end{table}
\begin{figure}[htb]
\centerline{
\epsfysize=3in
\epsffile{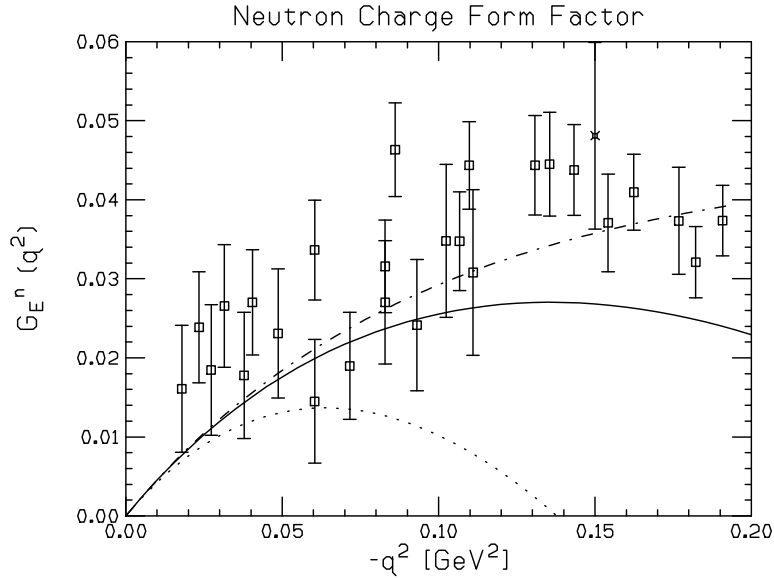}
}
\caption{The neutron electric form factor. The solid and the 
dotted lines are the third order SU(3) and SU(2) HBCHPT predictions,
respectively. The model--dependent data are taken from~\protect{\cite{pla}}
and the dot--dashed line is the dispersion--theoretical fit 
of~\protect{\cite{mmd}}. The filled cross indicates the
result of the Mainz $D(\vec{e},
e' \vec{n})$ experiment~\protect{\cite{mami}}.
}
\end{figure}

\medskip

Let us first discuss the implications for the nucleon sector (using
the results obtained with set~1). The effect of the kaon cloud on the
proton charge form factor is negligible, whereas the description
of the neutron charge form factor is obviously improved, as shown
in fig.1. This is mostly due to the fact that in SU(3), the kaon
cloud induces an additional momentum--dependence of the 
isoscalar electric and magnetic
form factors. In contrast, in SU(2) the isoscalar magnetic ff is
 simply a constant to third
order (as it was already pointed out in ref.\cite{hms}). Most likely, this 
sizeable
kaon cloud effect will be reduced when higher orders are calculated.
Also, the strangeness effects disimprove the description of the
magnetic form factors compared to the SU(2) results. As already stressed,
the magnetic coupling of the photon only starts at second order
and one thus should go to fourth order for a precise calculation of
the magnetic form factors (this is explicitely demonstrated for the 
magnetic moments in~\cite{mm}). It is also known from the SU(2)
calculation in ref.~\cite{bfhm} that the contribution from intermediate
decuplet states brings the magnetic radii closer to their empirical
values. In HBCHPT, such effects only start to appear at fourth order.

\begin{figure}
\epsfysize=2.5in
\epsffile{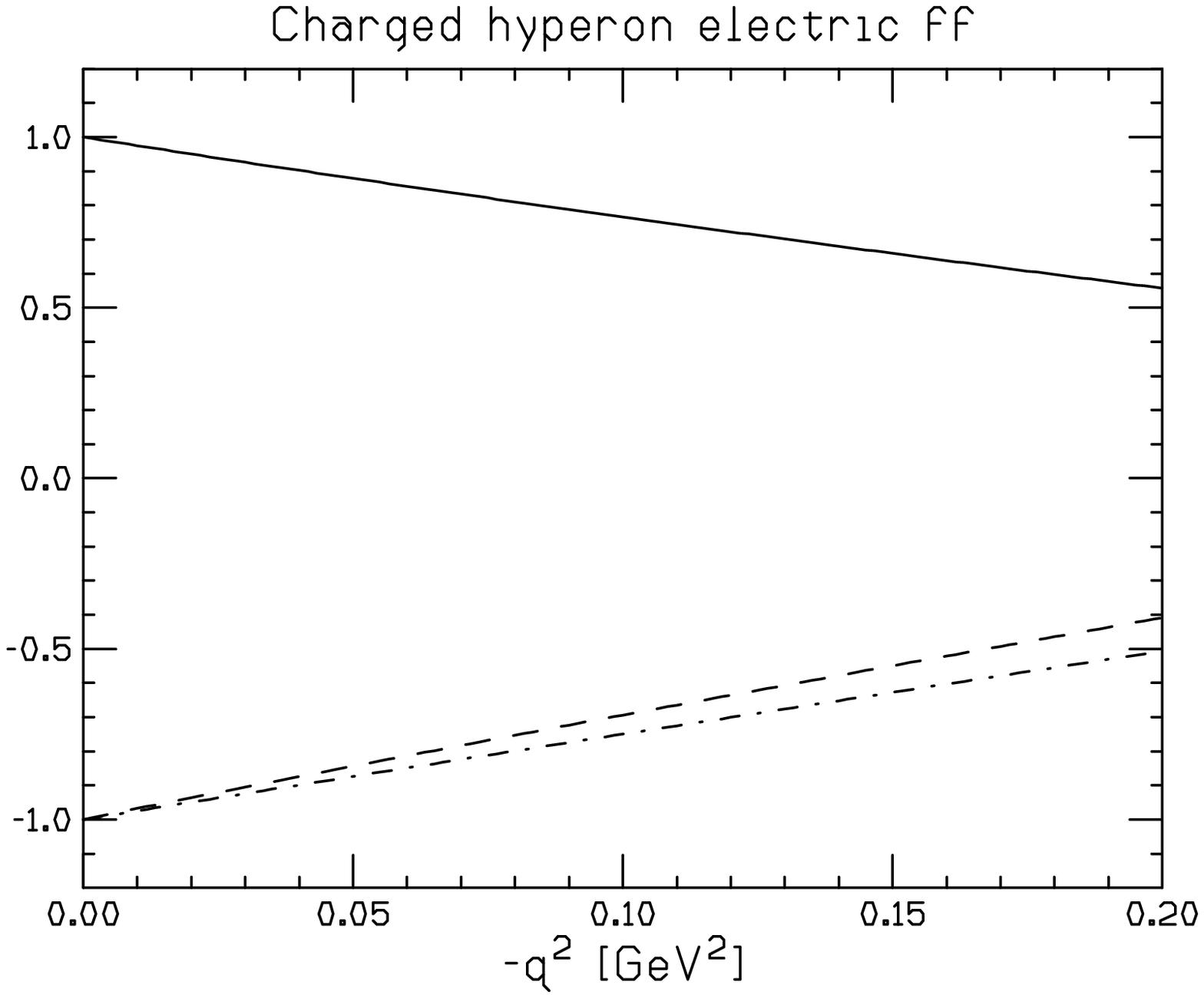}
\vspace{-2.5truein}
\hspace{3.3truein}
\epsfysize=2.5in
\epsffile{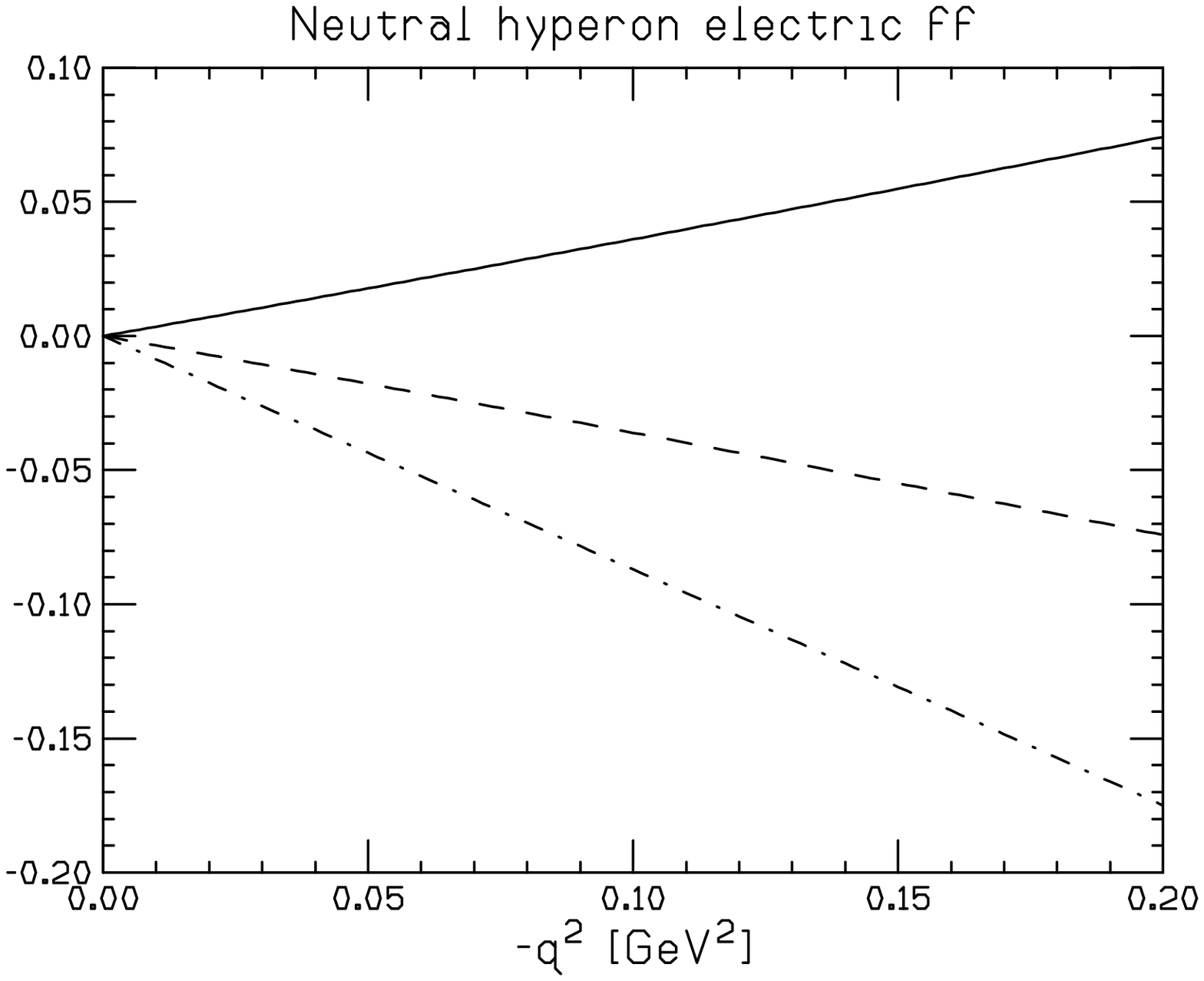}
\vskip 0.005cm
\caption{Electric hyperon form factors. Left panel: Charged particles,
$\Sigma^+$ (solid line), $\Sigma^-$ (dashed line), $\Xi^-$ (dot--dashed line).
Right panel: Neutral particles, $\Sigma^0$ (solid line), $\Lambda$ 
(dashed line), $\Xi^0$ (dot--dashed line).
}
\end{figure}

\begin{figure}
\epsfysize=2.5in
\epsffile{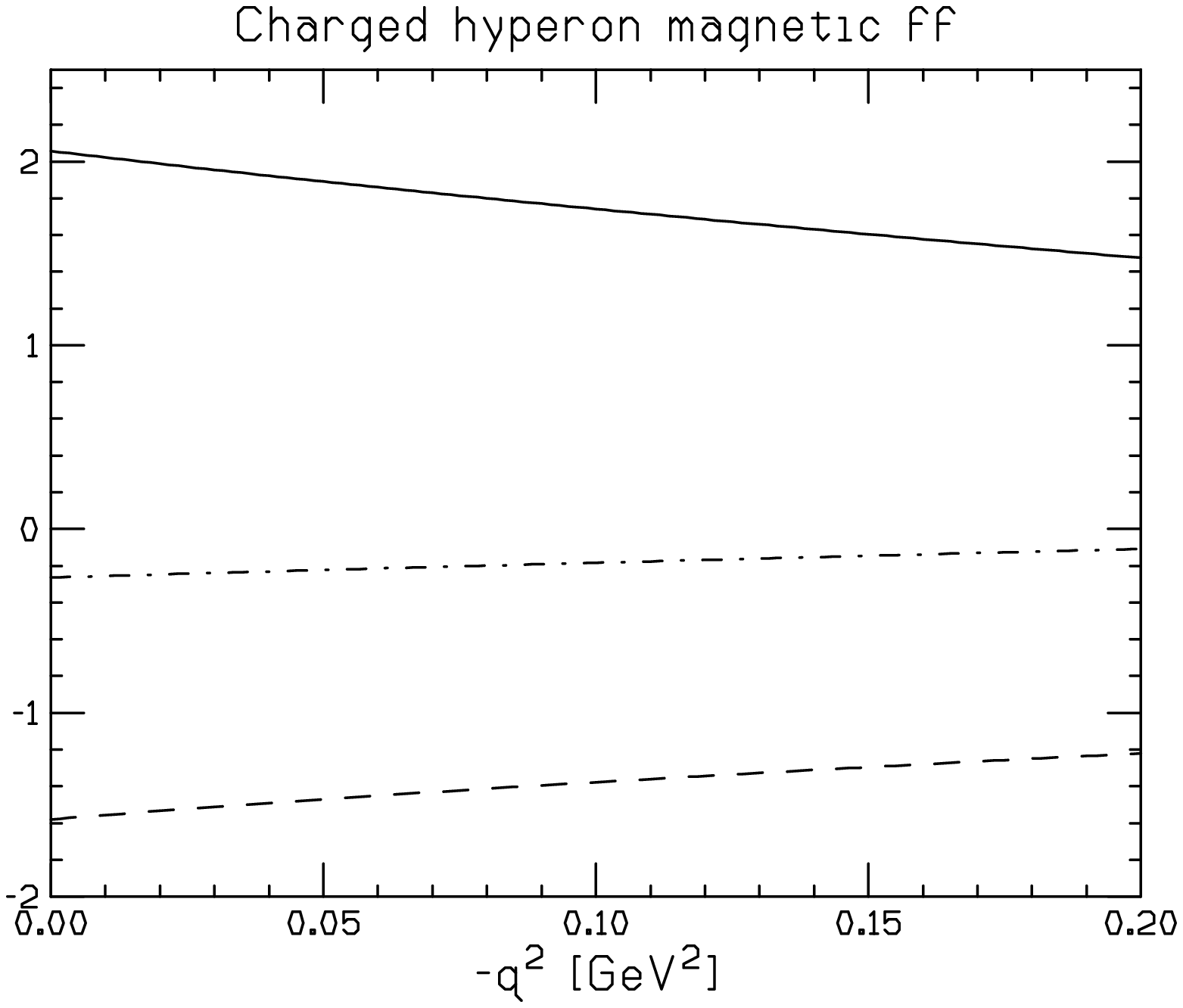}
\vspace{-2.5truein}
\hspace{3.3truein}
\epsfysize=2.5in
\epsffile{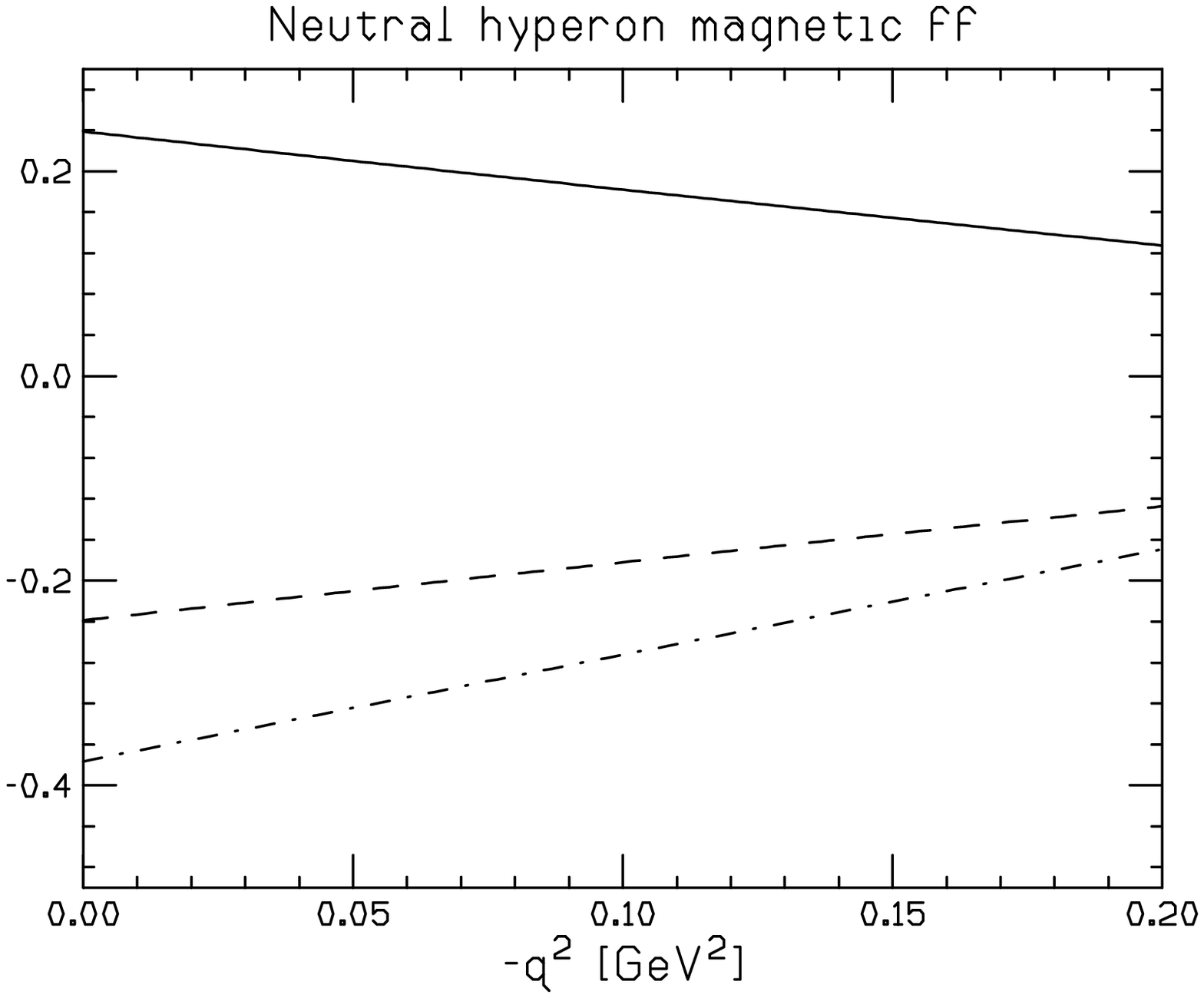}
\vskip 0.5cm
\caption{Magnetic hyperon form factors. Left panel: Charged particles,
$\Sigma^+$ (solid line), $\Sigma^-$ (dashed line), $\Xi^-$ (dot--dashed line).
Right panel: Neutral particles, $\Sigma^0$ (solid line), $\Lambda$ 
(dashed line), $\Xi^0$ (dot--dashed line).
}
\end{figure}

The charge radii and magnetic slopes of the hyperons can also be
found in table~2. The resulting numbers for the electric charge
radii are fairly stable under parameter variations for the charged
hyperons. We stress that the one for the $\Sigma^-$ agrees within
the quoted uncertainty with the measured values obtained at
CERN and FNAL. The corresponding electric and magnetic form factors
of all hyperons are collected in figs.~2,3 for parameter set~3.
It is worth to point out
that the electric ffs of the neutral hyperons do not show the
bending as the neutron electric ff does. This can be traced back
to the suppression of the pion cloud component for these particular
baryons. From the arguments given above, we expect the magnetic
slopes to be increased in magnitude when the calculation is done
to next order. The quality of the prediction for the magnetic form
factors is expected to be similar to the one for the magnetic moments.

\medskip

\noindent {\bf 5.} We have investigated the electromagnetic form factors
of the ground state hyperon octet
to leading one loop order in heavy baryon chiral perturbation theory.
To this order, one has four low--energy constants. Two of these can be
fixed from the magnetic moments and the other two from the electric
radii of the proton and the neutron. With that, the momentum--dependence
of the hyperon form factors is given parameter--free. The prediction 
for the $\Sigma^-$ radius agrees with the recently obtained value at
CERN and FNAL using high energetic hyperon beams. We have also derived
the so--called generalized Caldi--Pagels relations between various
charge radii. These are free of chiral loop effects and might
eventually lead to constraints on some neutral hyperon radii.
We also have investigated the
strangeness contribution to the nucleon electromagnetic form factors and
showed that strangeness can contribute significantly to the neutron charge
form factor. A next order calculation is called for to substantiate such
a claim.

\bigskip

\noindent {\bf Acknowledgements}

\medskip

\noindent
We thank Hans--Werner Hammer for a careful reading of the manuscript.
One of us (UGM) thanks the Institute for Nuclear Theory at
the University of Washington for its hospitality and the DOE for partial
support during the completion of this work.

\vspace{1truecm}

\end{document}